\documentclass[conference]{IEEEtran}
\IEEEoverridecommandlockouts
\usepackage{cite}
\usepackage{amsmath,amssymb,amsfonts}
\usepackage{algorithmic}
\usepackage{graphicx}
\usepackage{textcomp}
\usepackage{xcolor}
\def\BibTeX{{\rm B\kern-.05em{\sc i\kern-.025em b}\kern-.08em
    T\kern-.1667em\lower.7ex\hbox{E}\kern-.125emX}}
\begin{document}

\title{Unlocking High-Fidelity Analog Joint Source-Channel Coding on Standard Digital Transceivers
}

\author{Shumin Yao$^{1}$, Hao Chen$^{1}$, Yaping Sun$^{1}$, Nan Ma$^{2,1}$, \textit{Member, IEEE,} Xiaodong Xu$^{2,1}$, \textit{Senior Member, IEEE,} \\Qinglin Zhao$^{3}$, \textit{Senior Member, IEEE,} Shuguang Cui$^{4}$, \textit{Fellow IEEE}\\
$^{1}$Department of Broadband Communication, Pengcheng Laboratory, Shenzhen 518066, China \\
$^{2}$State Key Laboratory of Networking
and Switching Technology, Beijing University of Posts and \\Telecommunications, Beijing 100876, China\\
$^{3}$School of Computer Science and Engineering, Macau University of Science and Technology, \\Avenida Wei Long, Taipa, Macau, China\\
$^{4}$Future Nework of Intelligent Institute (FNii), \\ The Chinese University of Hong Kong (Shenzhen), Shenzhen, China \\
 \{yaoshm, chenh03, sunyp\}@pcl.ac.cn, \{manan, xuxiaodong\}@bupt.edu.cn, \\qlzhao@must.edu.mo, shuguangcui@cuhk.edu.cn
}

\maketitle

\begin{abstract}
Analog joint source-channel coding (JSCC) has demonstrated superior performance for semantic communications through graceful degradation across channel conditions. However, a fundamental hardware-software mismatch prevents deployment on modern digital physical layers (PHYs): analog JSCC generates continuous-valued symbols requiring infinite waveform diversity, while digital PHYs produce a finite set of discrete waveforms and employ non-differentiable operations that break end-to-end gradient flow. Existing solutions either fundamentally limit representation granularity or require impractical white-box PHY access. We introduce D2AJSCC, a novel framework enabling high-fidelity analog JSCC deployment on standard digital PHYs. Our approach exploits orthogonal frequency-division multiplexing's parallel subcarrier structure as a waveform synthesizer: computational PHY inversion determines input bitstreams that orchestrate subcarrier amplitudes and phases to emulate ideal analog waveforms. To enable end-to-end training despite non-differentiable PHY operations, we develop ProxyNet—a differentiable neural surrogate of the communication link that provides uninterrupted gradient flow while preventing JSCC degeneration. Simulation results for image transmission over WiFi PHY demonstrate that our system achieves near-ideal analog JSCC performance with graceful degradation across SNR conditions, while baselines exhibit cliff effects or catastrophic failures. By enabling next-generation semantic transmission on legacy infrastructure without hardware modification, our framework promotes sustainable network evolution and bridges the critical gap between analog JSCC's theoretical promise and practical deployment on ubiquitous digital hardware.
\end{abstract}


\section{Introduction}

Driven by rapid advancements in artificial intelligence, semantic communication has emerged as a transformative paradigm in wireless communications~\cite{SemComSurvey}. A key enabling technology is deep-learning-based joint source-channel coding (JSCC)~\cite{JSCCPractical}, where a deep neural network (DNN) encoder extracts essential semantic information from source messages while a corresponding decoder reconstructs them. By jointly optimizing this end-to-end process, JSCC robustly delivers information over noisy channels with minimal overhead, surpassing traditional separate source and channel coding schemes.

Among various JSCC approaches, analog JSCC has proven particularly effective~\cite{10614642}. It compresses high-dimensional sources into compact, continuous-valued symbols that capture fine-grained semantic nuances and are optimized for channel robustness. Pioneering work demonstrated significant performance gains for image transmission~\cite{bourtsoulatze2019deep}, spurring extensions to various modalities and scenarios. The hallmark of analog JSCC is \textit{graceful degradation}: performance smoothly scales with channel quality rather than exhibiting the catastrophic ``cliff effect" of digital systems.

However, a critical gap exists between analog JSCC's theoretical promise and practical deployment. These models assume an \textit{idealized analog-native physical layer (PHY)} that directly transmits continuous-valued waveforms. Modern hardware, however, is built on \textit{digital-centric PHYs} engineered to process discrete bit sequences. This hardware-software mismatch creates two fundamental deployment barriers.

\textit{Challenge 1: Signal format incompatibility.} Continuous JSCC symbols conceptually require infinite waveform diversity, but digital PHYs accept only bits and produce a finite set of discrete waveforms. The conventional workaround—quantizing symbols into bits—creates an intractable trade-off: high-precision quantization preserves fidelity but generates large payloads that negate JSCC's compression benefits, while low-precision quantization maintains efficiency but introduces catastrophic errors.

\textit{Challenge 2: Non-differentiable training barrier.} JSCC's power derives from end-to-end gradient-based optimization. Digital PHYs contain non-differentiable operations (channel coding, discrete modulation) that break gradient flow. Even if addressable via gradient approximations~\cite{jiang2022deep}, the PHY's internal channel coding forces the JSCC model to degenerate into a simple source coder, recreating separate source-channel coding and reintroducing the very cliff effect JSCC aims to eliminate.

Existing solutions remain inadequate. Quantization-based approaches~\cite{jiang2022deep, JSCCEngineering} still bottleneck performance through the fundamental information loss of discretization. Digital JSCC schemes attempt native compatibility but face their own limitations. Modulation-aware methods~\cite{DeepJSCCQ, JCM} require impractical ``white-box" PHY access to disable channel coding and directly control modulators—infeasible with integrated, black-box commercial hardware. Feature quantization approaches~\cite{D2JSCC, VQDeepSC} convert semantics to discrete codebooks, fundamentally limiting representation granularity.

This paper bridges the gap between analog JSCC's conceptual power and digital hardware's practical constraints. We introduce D2AJSCC, a novel framework enabling high-fidelity analog JSCC deployment on standard digital PHYs for the first time.  This approach not only unlocks the robustness of semantic communications but also enhances sustainability by repurposing existing digital infrastructure for analog-native protocols, thereby negating the need for costly hardware replacements. We demonstrate the practical viability of this framework by deploying an analog JSCC model for image transmission over a standard WiFi PHY. Our key contributions are:

\begin{itemize}
    \item To address the signal format incompatibility (Challenge 1), we introduce a waveform emulation approach that exploits the parallel subcarrier structure of orthogonal frequency-division multiplexing based (OFDM-based) digital PHYs as a flexible waveform synthesizer. Through computational PHY inversion, we determine the precise input bitstream that orchestrates subcarrier amplitudes and phases to collectively emulate the ideal analog JSCC waveform. This resolves the fidelity-efficiency impasse.
    
    \item To overcome the non-differentiable training barrier (Challenge 2), we develop a proxy-based training strategy. We first prevent JSCC degeneration by using software-defined modules (SDMs) to computationally invert the PHY's processing pipeline, rendering its channel coder transparent and compelling true joint optimization. We then introduce ProxyNet, a differentiable neural network trained as a high-fidelity surrogate of the entire SDM-enhanced communication link. During JSCC training, this frozen proxy creates an uninterrupted gradient path across non-differentiable hardware.
    
    \item We validate our framework via simulations of image transmission over standard 802.11a WiFi PHY. Results show our system nearly matches ideal analog JSCC performance with graceful degradation across SNR conditions, while quantization baselines exhibit cliff effects or catastrophic failures.
\end{itemize}

The remainder of this paper presents our system model and problem formulation (Sec.~\ref{sec:prelim}), the proposed D2AJSCC framework (Sec.~\ref{sec:framework}), experimental validation (Sec.~\ref{sec:eval}), and conclusions (Sec.~\ref{sec:conclusion}).

\section{System Model and Problem Formulation}
\label{sec:prelim}

This section provides the necessary background on analog JSCC and the WiFi PHY, then formulates the core deployment challenges our framework addresses.

\subsection{Analog Joint Source-Channel Coding}

\begin{figure}
    \centering
    \includegraphics[width=0.85\linewidth]{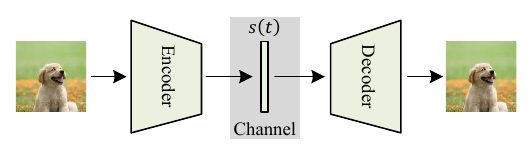}
    \caption{A typical analog joint source-channel coding system for image transmission.}
    \label{fig:AnalogJSCC}
\end{figure}

Analog JSCC jointly optimizes source and channel coding through deep neural networks, enabling graceful degradation where reconstruction quality degrades smoothly with channel conditions rather than exhibiting catastrophic cliff effects. As illustrated in Fig.~\ref{fig:AnalogJSCC}, a DNN encoder $f_{\text{enc}}$ compresses a source image $\mathbf{I} \in \mathbb{R}^{H \times W \times C}$ into a compact real-valued representation $\mathbf{z} \in \mathbb{R}^{2K \times 1}$:
\begin{equation}
    \mathbf{z} = f_{\text{enc}}(\mathbf{I}; \theta_{\text{enc}}).
\end{equation}
These features are mapped to $K$ complex symbols $\mathbf{S} \in \mathbb{C}^{K \times 1}$ by pairing consecutive elements:
\begin{equation}
    S_k = z_{2k-1} + j \cdot z_{2k}, \quad k=1, \ldots, K.
\end{equation}
The symbols are pulse-shaped into a continuous baseband waveform for transmission:
\begin{equation}
    s(t) = \sum_{k=1}^{K} S_k \cdot p(t - kT_s),
    \label{eq:Waveform}
\end{equation}
where $p(t)$ is a pulse-shaping filter and $T_s$ is the symbol period. This direct symbol-to-waveform mapping is crucial and assumes an analog-native PHY capable of transmitting continuous-valued signals.

At the receiver, the distorted symbols $\hat{\mathbf{S}}$ are decomposed back into real values $\hat{\mathbf{z}} \in \mathbb{R}^{2K \times 1}$ and fed to the decoder for reconstruction:
\begin{equation}
    \hat{\mathbf{I}} = f_{\text{dec}}(\hat{\mathbf{z}}; \theta_{\text{dec}}).
\end{equation}
The system is trained end-to-end by minimizing reconstruction error, typically mean squared error:
\begin{equation}
    \label{eq:JSCCLoss}
    \mathcal{L}_{\text{JSCC}}(\theta_{\text{enc}}, \theta_{\text{dec}}) = \left\| \mathbf{I} - \hat{\mathbf{I}} \right\|_2^2.
\end{equation}

\subsection{WiFi Physical Layer}

\begin{figure}
    \centering
    \includegraphics[width=1\linewidth]{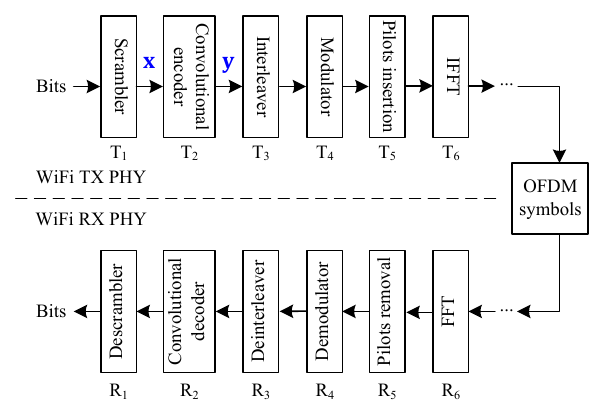}
    \caption{WiFi transmitter and receiver processing chains.}
    \label{fig:WiFiTXRX}
\end{figure}

The WiFi PHY operates on discrete bitstreams through a fixed, non-differentiable processing chain, as shown in Fig.~\ref{fig:WiFiTXRX}. The transmitter transforms input bits $\mathbf{b}$ through sequential modules: scrambling ($\mathrm{T}_{1}$), convolutional coding ($\mathrm{T}_{2}$), interleaving ($\mathrm{T}_{3}$), quadrature amplitude modulation (QAM, $\mathrm{T}_{4}$), pilot insertion ($\mathrm{T}_{5}$), and IFFT ($\mathrm{T}_{6}$) to generate an OFDM waveform. The receiver inverts this process through FFT ($\mathrm{R}_{1}$), channel equalization ($\mathrm{R}_{2}$), QAM demodulation ($\mathrm{R}_{3}$), deinterleaving ($\mathrm{R}_{4}$), convolutional decoding ($\mathrm{R}_{5}$), and descrambling ($\mathrm{R}_{6}$) to recover $\hat{\mathbf{b}}$.

\subsection{Problem Formulation} \label{sec:Challenges}

Deploying analog JSCC over a standard WiFi PHY presents two fundamental challenges stemming from the mismatch between software-defined analog systems and hardware-based digital architectures:

\textit{Challenge 1: Signal format incompatibility.} Analog JSCC generates continuous-valued symbols requiring infinite waveform diversity, while the WiFi PHY accepts only discrete bits and produces a finite set of waveforms. Quantizing JSCC symbols into bits creates an intractable trade-off: high-precision quantization preserves fidelity but generates large payloads that destroy transmission efficiency, while low-precision quantization maintains efficiency but introduces catastrophic quantization errors before transmission even begins.

\textit{Challenge 2: Non-differentiable training barrier.} JSCC's performance depends on end-to-end gradient-based optimization. The WiFi PHY contains non-differentiable operations that break gradient flow, preventing training. More critically, even if gradients could be approximated, the PHY's internal channel coding ($\mathrm{T}_{2}$) forces JSCC to degenerate into a simple source coder, recreating separate source-channel coding and reintroducing the cliff effect that JSCC aims to eliminate.

Our objective is to design a framework that enables high-fidelity analog JSCC deployment over standard WiFi PHY by resolving both challenges while preserving graceful degradation and transmission efficiency.

\section{Proposed Framework}
\label{sec:framework}

We introduce a two-component framework that enables analog JSCC deployment over standard WiFi PHY while maintaining both high fidelity and transmission efficiency.

\subsection{Framework Overview}

\begin{figure}
    \centering
    \includegraphics[width=1\linewidth]{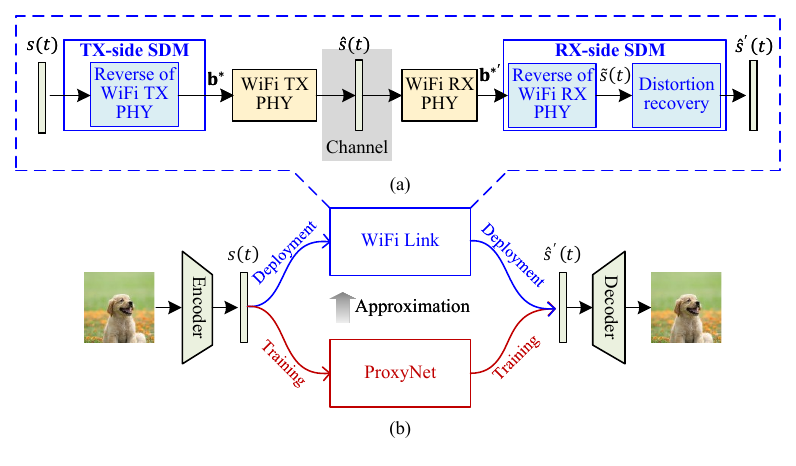}
    \caption{Framework overview: (a) Waveform emulation via PHY inversion using software-defined modules (SDMs). (b) ProxyNet enables end-to-end training through the non-differentiable physical link.}
    \label{fig:Overview}
\end{figure}

Our framework addresses the dual challenges via two core innovations, as illustrated in Fig.~\ref{fig:Overview}. First, we enable the digital WiFi PHY to emulate analog waveform transmission via PHY-layer inversion. Second, we enable gradient-based training through a differentiable proxy network that simulates the complete communication link.

\subsection{Waveform Emulation via PHY Inversion}

Our key insight is that a WiFi OFDM signal is a composite waveform constructed from many orthogonal subcarriers. By carefully selecting input bits to the WiFi TX PHY, we can control each subcarrier's amplitude and phase, orchestrating them to collectively approximate the target analog JSCC waveform $s(t)$. This is achieved by two SDMs that perform PHY-layer inversion.

As shown in Fig.~\ref{fig:Overview}(a), the sender-side SDM takes the target waveform $s(t)$ as input and computationally reverses the WiFi TX operations (from $\mathrm{T}_{6}$ back to $\mathrm{T}_{1}$) to find a specific bit sequence $\mathbf{b}^{*}$ that, when fed to the standard WiFi TX PHY, produces an output waveform $\hat{s}(t)$ approximating $s(t)$. The receiver-side SDM reverses the WiFi RX PHY process (from $\mathrm{R}_{1}$ back to $\mathrm{R}_{6}$) and employs a DNN-based recovery model to compensate for distortion, ensuring high-fidelity input to the JSCC decoder.

\subsubsection{Overcoming Non-Invertible Channel Coding}

The WiFi convolutional encoder ($\mathrm{T}_{2}$) poses a critical challenge: it performs a many-to-one mapping that is fundamentally non-invertible. For a desired output $\mathbf{y}$, a corresponding input $\mathbf{x}$ is not guaranteed to exist, potentially breaking the entire inversion chain.

We model the channel coding as a linear transformation over GF(2):
\begin{equation}
    \label{eq:ConvMatrix}
    \mathbf{C}_{\alpha \times \beta} \cdot \mathbf{x}_{\beta \times 1} = \mathbf{y}_{\alpha \times 1}
\end{equation}
where $\beta = R \cdot N \cdot \log_{2}M$ and $\alpha = N \cdot \log_{2}M$ for $N$ data subcarriers, $M$-QAM modulation, and coding rate $R < 1$. Since $R < 1$, the matrix $\mathbf{C}$ is "tall" ($\alpha > \beta$) and does not have full rank, making arbitrary $\mathbf{y}$ potentially unsolvable.

\begin{figure}
    \centering
    \includegraphics[width=0.85\linewidth]{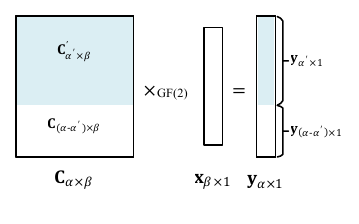}
    \caption{Solving non-invertibility by constraining to a subset of subcarriers, creating a square, full-rank system.}
    \label{fig:ConvMatrix}
\end{figure}

Our solution transforms this impossibility into guaranteed solvability by constraining emulation to a subset $N' < N$ of data subcarriers. This reduces the problem to:
\begin{equation}
    \label{eq:SubConvMatrix}
    \mathbf{C}'_{\alpha' \times \beta} \cdot \mathbf{x}_{\beta \times 1} = \mathbf{y}'_{\alpha' \times 1}
\end{equation}
where $\mathbf{C}'$ contains only the rows corresponding to our chosen subcarriers and $\alpha' = N' \cdot \log_{2}M$ (see Fig.~\ref{fig:ConvMatrix}). This system has a solution for any $\mathbf{y}'$ if $\mathbf{C}'$ has full row-rank, which is guaranteed when $\alpha' \leq \beta$. This yields the constraint:
\begin{equation} 
\label{eq:SubcrrBound}
    N' \leq R \cdot N
\end{equation}

By adhering to this bound, we ensure a valid $\mathbf{x}$ always exists. The unused $(N - N')$ subcarriers become ``dummy" carriers that the receiver-side SDM discards during reconstruction. This converts non-invertibility into a manageable trade-off between emulation bandwidth and coding rate.

\subsubsection{Distortion Compensation}

Even with successful inversion, the waveform suffers inevitable distortion from three sources: (1) quantization error from mapping continuous symbols to discrete QAM constellations, (2) cyclic prefix insertion that overwrites portions of the target waveform, and (3) uncontrollable pilot subcarriers with fixed values. These distortions share a key property: they exhibit multi-scale periodicity synchronized with both the WiFi OFDM symbol period and the JSCC symbol period.

We exploit this structure through a DNN-based compensation network inspired by TimesNet~\cite{wu2022timesnet}. The network transforms the 1D received waveform $\tilde{\mathbf{s}}_{\text{1D}} \in \mathbb{R}^{N_s \times 2}$ into two 2D representations that expose the periodic patterns:
\begin{equation}
    \label{eq:Reshape}
    \tilde{\mathbf{s}}_{\text{2D}}^{i} = \text{Reshape}_{i}(\text{Padding}(\tilde{\mathbf{s}}_{\text{1D}})), \quad i \in \{O, J\}
\end{equation}
where $O$ denotes OFDM periodicity and $J$ denotes JSCC periodicity. A shared 2D convolutional network $f_{\text{conv}}(\cdot; \theta_{\text{comp}})$ processes both representations, and the outputs are aggregated:
\begin{equation}
    \label{eq:Aggregate}
    \mathbf{s}_{\text{1D}}^{'} = \sum_{i \in \{O, J\}} \text{Trunc}(\text{Reshape}^{-1}(f_{\text{conv}}(\tilde{\mathbf{s}}_{\text{2D}}^{i}; \theta_{\text{comp}})))
\end{equation}

The network is trained to minimize mean square error (MSE) between recovered and ground-truth waveforms:
\begin{equation}
    \label{eq:CompObj}
    \theta_{\text{comp}}^* = \arg\min_{\theta_{\text{comp}}} \left\| \mathbf{s}_{\text{1D}}^{'} - \mathbf{s}_{\text{1D}} \right\|_2^2
\end{equation}

\subsection{ProxyNet for End-to-End Training}

The WiFi PHY's non-differentiable operations prevent direct end-to-end training of the JSCC model. Our solution is ProxyNet, a differentiable neural network trained to simulate the complete SDM-enhanced WiFi link behavior.

\subsubsection{Architecture}

ProxyNet consists of three sequential modules: a sender net $f_{\text{send}}(\cdot; \theta_{\text{send}})$ that mimics the sender-side SDM and WiFi TX PHY, a parameter-free but differentiable channel net that injects AWGN, and a receiver net $f_{\text{recv}}(\cdot; \theta_{\text{recv}})$ that models the WiFi RX PHY, receiver-side SDM, and distortion compensation. Both sender and receiver nets use U-Net architectures. The complete forward pass is:
\begin{equation}
    \begin{aligned}
        \hat{\mathbf{s}}_{\text{1D}} & = g_{\text{proxy}}(\mathbf{s}_{\text{1D}}; \theta_{\text{send}}, \theta_{\text{recv}}) \\ 
        &= f_{\text{recv}}(f_{\text{send}}(\mathbf{s}_{\text{1D}}; \theta_{\text{send}}) + \mathbf{n}; \theta_{\text{recv}})
    \end{aligned}
\end{equation}

ProxyNet is trained on input-output pairs from the real SDM-enhanced link:
\begin{equation}
    \label{eq:ProxyObj}
    (\theta_{\text{send}}^*, \theta_{\text{recv}}^*) = \arg\min_{\theta_{\text{send}}, \theta_{\text{recv}}} \left\| \hat{\mathbf{s}}_{\text{1D}} - \mathbf{s'}_{\text{1D}} \right\|_2^2
\end{equation}

\subsubsection{Three-Stage Training Strategy}

We employ a multi-stage approach to jointly optimize the JSCC model, distortion compensation network, and ProxyNet:

Stage 1: Distortion compensation initialization. We train the compensation network by transmitting known waveforms through the link and minimizing:
\begin{equation}
    \mathcal{L}_{\text{comp}} = \left\| \mathbf{s'}_{\text{1D}} - \mathbf{s}_{\text{1D}} \right\|_2^2
\end{equation}

Stage 2: ProxyNet initialization. Using the pre-trained compensation network, we train ProxyNet on collected link input-output pairs:
\begin{equation}
    \mathcal{L}_{\text{proxy}} = \left\| \hat{\mathbf{s}}_{\text{1D}} - \mathbf{s'}_{\text{1D}} \right\|_2^2
\end{equation}

Stage 3: Alternating joint optimization. We iteratively alternate between two phases until convergence. First, with the frozen ProxyNet, we train the JSCC model and fine-tune the compensation network end-to-end by minimizing:
\begin{equation}
    \mathcal{L}_{\text{total}} = \mathcal{L}_{\text{JSCC}} + \gamma \mathcal{L}_{\text{comp}}
\end{equation} where $\mathcal{L}_{\text{JSCC}} = \| \mathbf{I} - \hat{\mathbf{I}} \|_2^2$ is the image reconstruction loss and $\gamma \in [0,1]$ balances the objectives. Second, with frozen JSCC and compensation networks, we generate new waveforms using the updated encoder, transmit them through the real link, and fine-tune ProxyNet on these fresh samples to maintain accurate link simulation. This alternating strategy ensures ProxyNet continuously adapts to the evolving JSCC model while enabling true end-to-end gradient flow for joint optimization.

\begin{figure}
    \centering
    \includegraphics[width=0.85\linewidth]{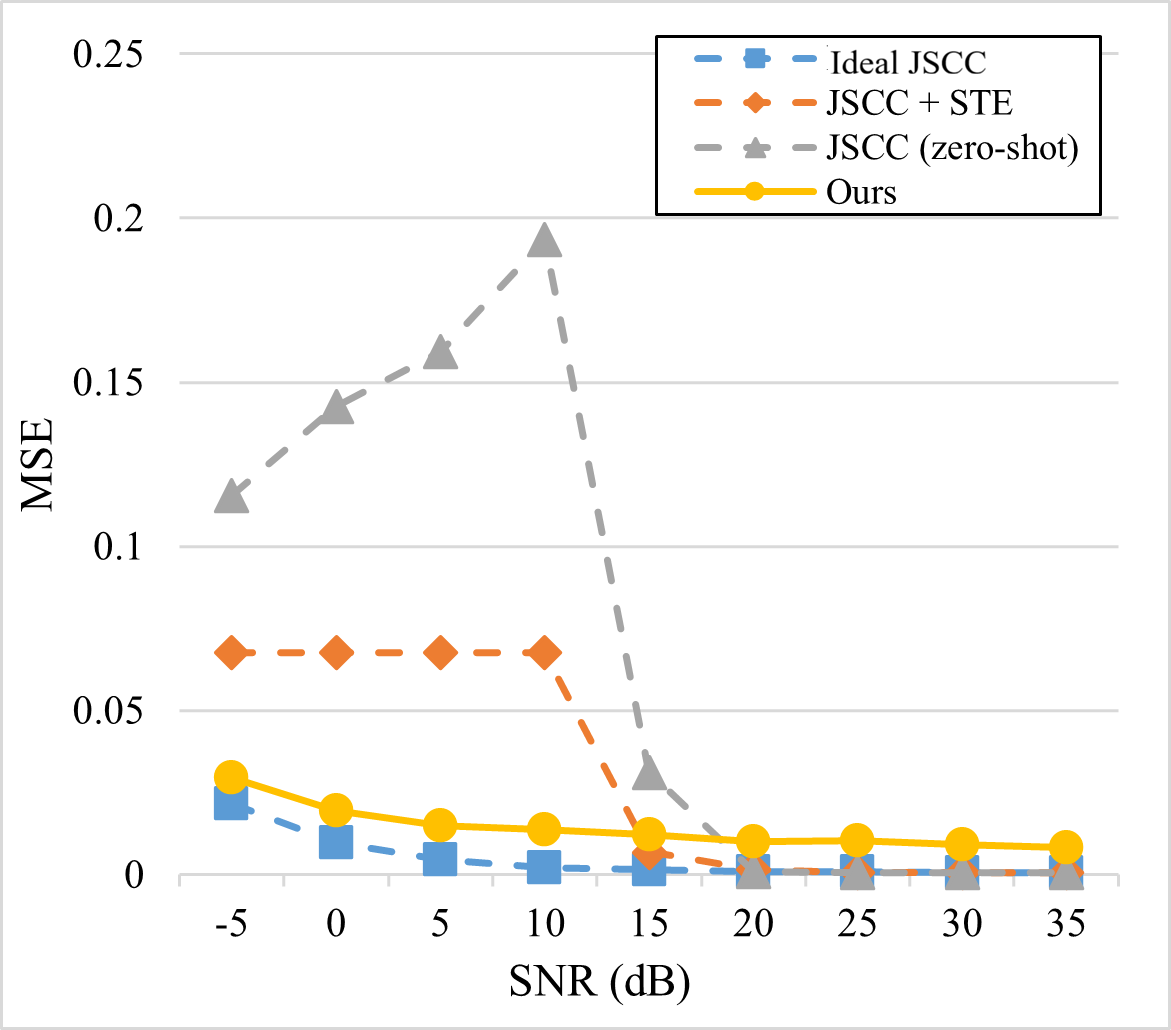}
    \caption{MSE between source and reconstructed images versus SNR.}
    \label{fig:MSE}
\end{figure}

\begin{figure*}[t!]
    \centering
    \includegraphics[width=0.85\linewidth]{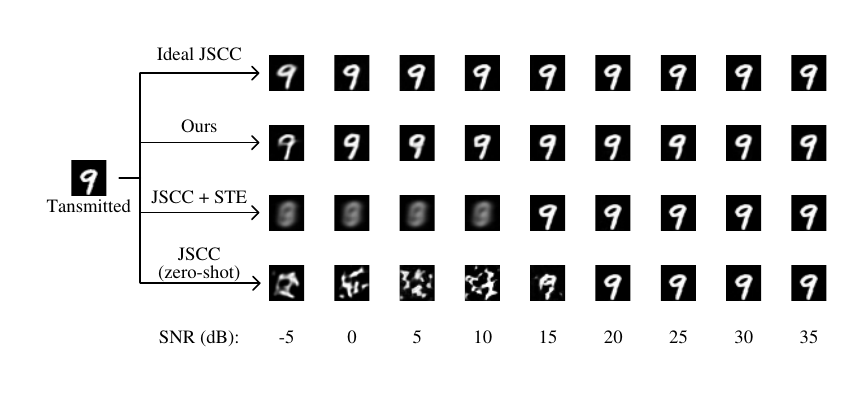}
    \caption{Visual comparison of reconstructed images across SNR range from -5 dB to 35 dB.}
    \label{fig:RXVisual}
\end{figure*}

\section{Experimental Results}\label{sec:eval}

\subsection{Experimental Setup}

We evaluate our framework on semantic image transmission over a WiFi link. Using the MNIST dataset as a proof-of-concept benchmark, we implement the DNN components (JSCC models and ProxyNet) in PyTorch and train on an NVIDIA A6000 GPU. The SDM-enhanced WiFi link simulates IEEE 802.11a non-HT transmission over a 20 MHz channel using MCS 7 (64-QAM with 5/6 coding rate), implemented in MATLAB using WLAN toolboxes.

We compare against three baselines: (1) \textit{Ideal JSCC} assumes a perfect analog PHY that directly transmits continuous-valued waveforms, representing the unattainable performance ceiling free from hardware-software mismatch. (2) \textit{JSCC + STE} trains the JSCC model end-to-end over the WiFi link by serializing floating-point encoder outputs into bits, using the straight-through estimator (STE) for gradient approximation through non-differentiable PHY operations. (3) \textit{JSCC (zero-shot)} deploys an AWGN-trained JSCC model directly on the WiFi link without adaptation, quantifying the severe degradation caused by hardware-software mismatch.

\subsection{Performance Evaluation}

Fig.~\ref{fig:MSE} plots the MSE between source and reconstructed images over SNR ranging from -5 dB to 35 dB. The results reveal several critical insights. First, both ``Ideal JSCC'' and our method exhibit graceful degradation as SNR decreases—the hallmark of true JSCC systems. Our MSE curve closely tracks the ideal baseline across the full SNR range, with the small gap attributable to unavoidable hardware-induced distortions (quantization error, cyclic prefix mismatch, pilot interference) inherent to the WiFi PHY. This near-ideal performance validates our framework's effectiveness in bringing analog JSCC benefits to practical digital implementations.

Second, the ``JSCC + STE'' baseline suffers from the classic ``cliff effect,'' maintaining high MSE below 15 dB SNR before dropping sharply. This validates the channel coding degeneracy problem described in Section~\ref{sec:Challenges}: the WiFi PHY's channel coding forces the JSCC model to degenerate into simple source coding, recreating a separate source-channel system whose performance is dictated entirely by the channel coding's operating threshold. Below this threshold, reconstruction fails catastrophically. Our method successfully avoids this separation.

Third, the ``JSCC (zero-shot)'' baseline demonstrates catastrophic failure due to hardware-software mismatch. In the low-to-mid SNR range (up to 10 dB), its MSE is extremely high and counter-intuitively worsens as SNR improves. The AWGN-trained model misinterprets the structural WiFi PHY distortions as overwhelming noise; as actual channel noise decreases with rising SNR, these un-learned hardware artifacts become dominant. The sharp performance improvement when SNR $>$ 10 dB reflects the WiFi PHY's channel coding establishing a near-perfect digital link, not JSCC model success.

Fig.~\ref{fig:RXVisual} provides qualitative validation through visual comparison of reconstructed images across the SNR range. Both our method and ``Ideal JSCC'' demonstrate smooth, progressive quality improvement as SNR increases. At low SNRs (e.g., -5 dB), images remain blurry but retain coarse structural shape; as SNR improves, digits become increasingly clear and recognizable. This exemplifies well-functioning JSCC, where information quality gracefully scales with channel quality. Our visual quality closely tracks the ideal baseline at any given SNR.

The ``JSCC + STE'' baseline exhibits a hard cliff effect, failing to reconstruct recognizable images below 15 dB SNR—outputs are merely blurry blobs. At the 15 dB threshold, quality suddenly improves to clear digits, confirming that performance is dictated by the underlying WiFi channel coding rather than joint optimization.

The ``JSCC (zero-shot)'' baseline provides the most dramatic illustration of hardware-software mismatch. Its reconstruction quality catastrophically collapses as SNR increases from -5 dB to 10 dB. Instead of becoming clearer, digit structure disintegrates into disconnected, high-contrast splotches. This counter-intuitive collapse occurs because the AWGN-trained model misinterprets structured WiFi PHY distortion as noise. As actual channel noise decreases, these un-modeled artifacts dominate and the decoder fails completely. Quality only recovers for SNR $>$ 10 dB when WiFi channel coding establishes a near-error-free link—not because the JSCC model functions correctly.

These results demonstrate that our framework successfully emulates classic analog JSCC systems' graceful degradation while operating on entirely digital hardware. We make a strategic trade-off: accepting a minor performance ceiling at high SNR (due to residual emulation error) in exchange for unparalleled robustness in noisy, variable environments. This inherent resilience makes our method fundamentally more reliable and practical for real-world communication systems than conventional digital approaches optimized for narrow, ideal conditions.

\section*{Acknowledgment}

This work is supported by the National Science and Technology Major Project - Mobile Information Networks under Grant No. 2024ZD1300700, and by National Natural Science Foundation of China with grant 62301471.

\section{Conclusion}\label{sec:conclusion}

This paper introduced D2AJSCC, a novel framework that bridges the critical gap between analog JSCC's demonstrated advantages and practical deployment on ubiquitous digital hardware. By exploiting OFDM's parallel subcarrier structure as a waveform synthesizer, our computational PHY inversion approach enables high-fidelity emulation of continuous-valued analog waveforms using standard digital PHYs. Our proxy-based training strategy, centered on ProxyNet as a differentiable neural surrogate, enables true end-to-end gradient flow while preventing JSCC degeneration caused by non-differentiable PHY operations. Simulation results demonstrate that our system achieves near-ideal analog JSCC performance with graceful degradation across SNR conditions, successfully avoiding the cliff effects that plague conventional digital approaches. This work establishes a practical pathway for deploying the superior robustness and efficiency of analog JSCC on modern communication hardware, fostering a sustainable ecosystem where advanced semantic capabilities coexist with legacy infrastructure.


\bibliographystyle{ieeetr}
\bibliography{ref}

\end{document}